\documentclass[aps,prb,reprint,superscriptaddress,floatfix,10pt,twocolumn,eprint]{revtex4-2}
\usepackage{caption}
\usepackage{ragged2e}
\usepackage{subcaption}
\usepackage{comment}
\usepackage{graphicx}
\usepackage{dcolumn}
\usepackage{amsmath,amssymb,amsfonts}
\usepackage{latexsym,verbatim}
\usepackage{bm}
\usepackage{color}
\usepackage{cancel}
\usepackage{multirow}
\usepackage[abs]{overpic}
\usepackage[colorlinks=true,citecolor=blue,linkcolor=blue,urlcolor=blue]{hyperref}
\usepackage{overpic}
\usepackage{array}
\usepackage{nicefrac}
\usepackage{physics}
\usepackage{bbold}
\usepackage{soul}
\usepackage{enumerate}
\usepackage{xcolor}
\usepackage{upgreek} 
\usepackage{braket}
\usepackage{transparent}
\usepackage{tikz}

  \newcommand{\vxi}{\vec{\xi}}
  \newcommand{\MM}{\mathbf{M}}
		\newcommand{\MLambda}{\mathbf{\Lambda}}
\begin{document}

\title{Enhanced thermoelectric effects in a driven one-dimensional system}

\author{C. X. Zhang}
\affiliation{NEST, Istituto Nanoscienze-CNR and Scuola Normale Superiore, I-56126 Pisa, Italy}

\author{Alessandro Braggio}
\affiliation{NEST, Istituto Nanoscienze-CNR and Scuola Normale Superiore, I-56126 Pisa, Italy}
\affiliation{Institute for Quantum Studies, Chapman University, Orange, CA 92866, USA
}
\author{Alessandro Romito}
\affiliation{Department of Physics, Lancaster University, Lancaster LA1 4YB, United Kingdom}

\author{Fabio Taddei}
\affiliation{NEST, Istituto Nanoscienze-CNR and Scuola Normale Superiore, I-56126 Pisa, Italy}

\begin{abstract}
We investigate the thermoelectric properties of a one-dimensional quantum system in the presence of an external driving. We employ Floquet scattering theory to calculate linear-response stationary thermoelectric figures of merit in a single-channel conductor subjected to a periodically varying delta-like potential barrier. We also include a step barrier in one of the leads as a model of a nanoscale inhomogeneous semiconducting system.
In the absence of a step barrier, we found that external driving can significantly enhance the Seebeck coefficient, particularly at low temperatures, with a relative increase of up to 200\% at high frequencies compared to the static case.
In the presence of a step barrier, we found that the thermoelectric Onsager coefficient for the driven case is also enhanced compared to the static case, with a significant photon-assisted effect at low temperatures when the chemical potential is within the semiconductor's gap.
Our results demonstrate that external driving can be used to tune and enhance the thermoelectric capabilities of low-electron-density nanodevices.
\end{abstract}

\maketitle

\section{Introduction}
Heat management is crucial for achieving high qubit density and optimal performance in quantum processors~\cite{Auffeves_2022}.
Thermoelectric effects provide us with a clever way to cool via optimally controlling heat flows and, at the same time, convert it into electricity~\cite{Lee_2016}, which can be reused at the processor level or easily piped to higher temperatures where more efficient cooling mechanisms can be employed~\cite{Antola_2024}.
In this perspective, the thermoelectric properties of
mesoscopic nanodevices~\cite{Benenti_2017} routinely employed in quantum technologies are of immediate interest. 
A paradigmatic case are thermoelectric effects in superconductors, which are at the core of several recently proposed thermoelectrically based quantum devices~\cite{Jones_1947,Vechten_1999,Giazotto_2015,Varpula_2017,Hekkila_2018,Paolucci_2023,Bergeret_2018,Marchegiani_2020a,Blasi_2020,Blasi_2020a,Marchegiani_2020b,Blasi_2021,Singh_2024,Germanese_2022,Arrachea_2025}.

Operating nano-devices through external periodic AC driving (or photon-assisted mechanisms) is a natural approach to heat management (for example, for the implementation of heat engines, refrigerators, and heat pumps)
and for improving their thermoelectric capabilities,
possibly taking advantage of quantum effects, see for example Refs.~\cite{Moskalets2002-2,Arrachea_2007_PhysRevB.75.245420, Moskalets_2009_PhysRevB.80.081302, Crepieux_2011_PhysRevB.83.153417, Juergens_2013_PhysRevB.87.245423, Lim_2013_PhysRevB.88.201304,Chen_Q_2013, Chen_XB_2013, Tagani_2014, Virtanen_2014, Chen_2015, Rossello2015,Ludovico2016,Bruch2016,Ludovico2016-2,Ludovico2016-3,Gallego-Marcos_2017,Molignini2017,Haughian2018,Erdman_2019,Brandner_2020,Sengupta_2020,Ganguly_2020,Bhandari2020,Cangemi2020,Potanina2021,Cavina2021,Izumida2021,Lu2022,Ryu_2022,Monsel2022,Hijano_2023,Slimane_2023,Lopez2024,Deghi2024,Chowdhury_2024,Lu2025,Aguilar2024,Acciai2025}.
The vast majority of such papers deal with the adiabatic regime, where the frequency of the periodic driving is the smallest energy scale, whereas only a few concern high frequencies, which lie beyond the adiabatic regime.
A photon-assisted refrigerator has been experimentally realized in Ref.~\cite{Kafanov_2009}.

From a modeling perspective, quantum coherent transport for non-interacting periodically driven systems can be fruitfully described using Floquet scattering theory~\cite{Moskalets2002, Moskalets2004, Moskalets2008, Stefanucci2008}, which can be considered an analogue of the Bloch theory in the time domain.

In this paper, we investigate the effect of AC driving on the stationary linear thermoelectric current in one-dimensional, quantum-coherent, non-interacting electronic systems. Specifically, we employ the Floquet scattering formalism to calculate the electric and thermoelectric Onsager coefficients for a system comprising a driven tunnel barrier situated between two leads. This analysis also accounts for an energy band shift between the leads, modeled as a step potential. We study the Onsager and Seebeck coefficients as functions of temperature and driving frequency, considering various parameters, including driving amplitude, the strength of the static barrier component, and the height of the step potential. In particular, we focus on the high-frequency regime, where the driving frequency can be comparable to other energy scales, such as the chemical potential, temperature, or energy gaps.
This regime is experimentally accessible in systems with sufficiently low electron density, such as semiconductors or two-dimensional materials, where electron density can be readily controlled by gating.
We find that driving can enhance the Seebeck coefficient, particularly at low temperatures, by increasing the strength of the AC component of the barrier. Similarly, in the presence of an energy band shift, we observe an enhancement of the thermoelectric Onsager coefficient compared to the corresponding static case.

The paper is structured as follows:
In Sec.~\ref{FlTr}, we discuss the general properties of stationary linear thermoelectric coefficients in the presence of AC driving, utilizing Floquet scattering theory. In Sec.~\ref{secMod}, we apply this theory to a one-dimensional conductor, detailing the computation of Floquet scattering matrices. Our results are presented in Sec.~\ref{secRes}, where we show how periodic external driving can enhance linear thermoelectric capabilities for both cases: with and without a step potential. Finally, in Sec.~\ref{secConclusion}, we conclude by discussing potential implications and outlining future research directions.

\section{Floquet transport}
\label{FlTr}
We consider a generic multi-terminal system subject to a time-dependent periodic driving with frequency $\omega$.
We are interested in the charge current $J_e^\alpha$ flowing in terminal $\alpha$ averaged over a period $T=2\pi/\omega$.
Using the Floquet scattering approach~\cite{Moskalets2002, Moskalets2004, Moskalets2008,Bruch_2018}, we can rewrite $J_e^\alpha$ as
\begin{align}
     J_{e}^\alpha 
     =\frac{e}{h}\!\!\int_{-\infty}^{\infty}\!\!\!\!\!\!\!\! dE \sum_{\beta,n}\,^{'}  \left| {\cal S}_{\alpha\beta}(E,E_n)\right|^2\left[    f_\beta(E_n)
    - f_\alpha(E)\right],
    \label{jea1}
\end{align}
where $E_n=E +n \hbar\omega$, with $n\in \mathbb{Z}$.
The integration over the energies and the summation $\sum_n^{'}$ are computed only for propagating states in the leads, i.e., $E>\epsilon_\alpha$ and $E_n>\epsilon_\beta$, where $\epsilon_i$ denotes the energy band bottom for $i$-lead.
In Eq.~(\ref{jea1}), ${\cal S}_{\alpha\beta}(E_2,E_1)$ is the scattering amplitude for an electron injected from lead $\beta$ toward the scattering region with
energy $E_1$ and arriving in lead $\alpha$ with energy $E_2$, and $f_\alpha(E)=\{1+\exp[(E-\mu_\alpha)/k_B T_\alpha]\}^{-1}$ is the Fermi function relative to lead $\alpha$, where $\mu_\alpha$ and $T_\alpha$ are, respectively, its temperature and chemical potential.
Thus
$\left| {\cal S}_{\alpha\beta}(E,E_n)\right|^2$ represents the probability for the transfer of an electron from lead $\beta$ to $\alpha$ with the emission of $n$ photons of energy $\hbar\omega$.
For the propagating states, one can show the unitarity conditions~\cite{Moskalets2002}
\begin{equation}
\sum_{n,\beta}\,^{'}
|{\cal S}_{\alpha\beta}(E,E_n)|^2
=\sum_{n,\alpha}\,^{'}  |{\cal S}_{\alpha\beta}(E_n,E)|^2=1\ .
\end{equation}

By properly applying unitarity and making a convenient change of variables in the energy integration, along with adjusting the sign of $n$, we can reduce the current in lead $\alpha$ to a more conventional form~\cite{Moskalets2002}:
\begin{align}
    J_e^\alpha=\frac{e}{h}\int_{-\infty}^{\infty} \!\!\!\!\!dE \sum_{\beta}\left[   \mathcal{T}_{\alpha\beta}(E) f_\beta(E)   - \mathcal{T}_{\beta\alpha}(E)  f_\alpha(E)\right] .\label{chcu}
\end{align}
Here we have introduced the transmission coefficients defined as
\begin{align}
\label{eq:Tau}
\mathcal{T}_{\alpha\beta}(E)=\sum_{n}\,^{'}
\left| {\cal S}_{\alpha\beta}(E_n,E)\right|^2,
\end{align}
where the sum must be computed only over the propagating states in the leads.
In the following, we will concentrate on two-terminal devices so that $J_e\equiv J_e^L=-J_e^R$ and the transmission coefficients are only $\mathcal{T}_{RL}(E)$ and $\mathcal{T}_{LR}(E)$.

Since we will investigate how the AC driving affects the thermoelectric quantities in linear response, 
it is useful to introduce the linear expansion of the current for a system at temperature $T$ in terms of the affinities. Specializing to a two-terminal device, we can write the affinities as $\mathcal{F}_e=\Delta V / T$ and $\mathcal{F}_h= \Delta T / T^2$ for small voltage bias $\Delta V\ll k_BT/e$ and temperature difference $\Delta T\ll T$.
We also assume that chemical potentials and temperatures are applied symmetrically, i.e., $\mu_{L/R}=\mu\pm e\Delta V/2$ and $T_{L/R}=T\pm\Delta T/2$.
Thus, the current can be written as
\begin{align}
\label{je}
    J_e = J_e^P + L_{\rm ee}\mathcal{F}_e + L_{\rm eh}\mathcal{F}_h,
\end{align}
where $J_e^P$ is the pumped current which is zero when $\mathcal{T}_{RL}(E)=\mathcal{T}_{LR}(E)$, see App.~\ref{Appump}.
The Onsager coefficients $L_{\rm ee}$ and $L_{\rm eh}$ can be directly connected to relevant physical quantities such as the two-terminal conductance $G=T^{-1}L_{\rm ee}$ and the Seebeck coefficient~\cite{Ashcroft+Mermin}
\begin{align}
S=\frac{1}{T}\frac{L_{\rm eh}}{L_{\rm ee}}.
\label{eq:Seebeck}
\end{align}
The latter quantity measures the strength of the thermoelectric effect, and is defined as the ratio $S=(\Delta V/\Delta T)_{J_e=0}$, where $\Delta V$ is the voltage necessary to stop the thermoelectric current generated by a temperature difference $\Delta T$.  

In particular, the Onsager coefficients in Eq.~(\ref{je}) can be expressed as
\begin{align}
L_{\rm ee}&=-\frac{e^2}{h}T\int dE \bar{\mathcal{T}} (E) f'(E) \label{Lee} ,
\end{align}
and
\begin{align}
L_{\rm eh}&=-\frac{e}{h}T\int dE (E-\mu)\bar{\mathcal{T}} (E) f'(E) ,\label{Leh}
\end{align}
where we define $\bar{\mathcal{T}} (E)=[\mathcal{T}_{RL}(E)+\mathcal{T}_{LR}(E)]/2$ and $f'(E)=\partial f/\partial E$ is the energy derivative of the Fermi function $f(E)$, whose chemical potential and temperature are $\mu$ and $T$, respectively.
This shows that the expressions of these two Onsager coefficients are formally equal to the ones relative to the static case,
where $\bar{\mathcal{T}} (E)$ is replaced by the transmission in the static case
~\cite{Benenti_2017}.
Due to the even symmetry in energy of the Fermi function derivative, $L_{\rm ee}$ is determined by the even-in-energy component of $\bar{\mathcal{T}}(E)$ around the Fermi energy, while $L_{\rm eh}$
measures the odd-in-energy symmetry of $\bar{\mathcal{T}} (E)$ around the Fermi energy. 

Furthermore, we note that for very small temperatures we can expand in energy the transmission function around the chemical potential such as $\bar{\mathcal{T}}(E)=\bar{\mathcal{T}}(\mu)+\bar{\mathcal{T}}'(\mu)(E-\mu)+\mathcal{O}[(E-\mu)^2]$.
Using the Sommerfeld expansion one can show that $L_{\rm ee}\simeq e^2 \bar{\mathcal{T}}(\mu) T/h$ and  $L_{\rm eh}\simeq (\pi^2/3)(e/h)k_B^2\bar{\mathcal{T}}'(\mu) T^3$~\cite{Benenti_2017}.
As a consequence, in the same limit, one finds that $G$ is independent of temperature, while  the Seebeck coefficient is linear in temperature, namely
\begin{equation}
S \simeq \left( \frac{k_B}{e}\right)
\left(\frac{\pi^2}{3}\right) \frac{\bar{\mathcal{T}}' (\mu)}{\bar{\mathcal{T}} (\mu)}k_B T ,
\label{Sapp}
\end{equation}
where the dimensions are given by the prefactor $k_B/e\simeq 81  \mu$V/K.

\section{Model}
\label{secMod}
\begin{figure}[ht]
\centering
\begin{tikzpicture}
\draw [->, thick] (-4,0) -- (4,0);
\node [above] at (3,0) {$x$};
\draw [->, thick] (0,-0.5) -- (0,3.5);
\node [above] at (-3.8,3) {$E$};
\node [above] at (-3.8,2) {$\mu$};
\node [above, text=blue] at (-0.7, 2.6) {$W(t)\delta(x)$};
\draw [<->, thick] (1,0) -- (1,1);
\node [above] at (1.3,0.5) {$\Delta$};
\draw[thick] (0,1) -- (4,1);
\draw[ultra thick, blue] (0,0) -- (0, 3.7);
\draw[thick, domain=-3.5 : -0.5] plot (\x, {(\x+2)^2});   
\draw[thick, domain=0.5 : 3.5] plot (\x, {1+(\x - 2)^2});   

\draw [dashed] (-4,2) -- (4,2);

\fill[gray!50] plot[domain=-3.41 : -0.59, samples=50] (\x, {(\x+2)^2}) 
-- plot[domain=-3.41 : -0.59, samples=2] (\x, {2}) 
-- cycle;

\fill[gray!50] plot[domain= 1 : 3, samples=50] (\x, {1+(\x - 2)^2}) 
-- plot[domain=1 : 3, samples=2] (\x, {2}) 
-- cycle;

\end{tikzpicture}
\caption{\justifying Sketch of the energy bands
of the 1D model where the blue line indicates the delta-function potential at $x=0$. The two parabolas represent the energy-momentum relation of electrons. If $\Delta > 0$, the bottom of the right band is shifted up with respect to the left band, as a sort of step-like barrier. The dashed line indicates the chemical potential, while the shaded regions represent the occupied states.
}
\label{fig:scheme}
\end{figure}
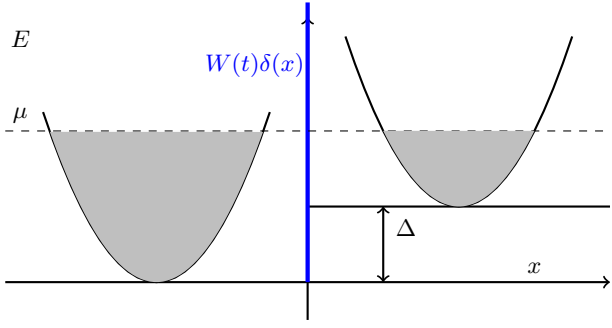

In this section, we apply the Floquet
approach to an ideal one-dimensional setup consisting of a single band electronic system with a driven time-periodic delta-like impurity.
The system is described by the following Hamiltonian 
\begin{equation}\label{ham}
H= \frac{\hbar^2 k^2}{2m} + W(t) \delta(x)
  +\Delta \theta (x),
\end{equation}
where $W(t)=W_0 + 2W_1 \cos(\omega t)$ is the time-dependent strength of a delta-like potential, and $\theta (x)$ is the Heaviside function.
Notice that the potential is a periodic function, i.e. $W(t+T)=W(t)$ with period $T= 2\pi/\omega$.
Moreover, we also include the possibility of an energy band shift $\Delta=\epsilon_R-\epsilon_L$, which represents the height of a band-mismatch located in $x=0$ ($\epsilon_R$ and $\epsilon_L$ are the energy band bottoms of the right lead and left lead, respectively, see Fig.~\ref{fig:scheme}).
Note that we consider a single driving parameter so that we cannot realize adiabatic pumping.

To calculate the scattering amplitudes appearing in the expressions for the currents, we use the fact that the potential $W(t)$ is periodic.
This allows us to apply the Floquet theorem 
\cite{Shirley_1965, Barone_1977, Martinez_2001, Moskalets2002}, so that the solution of the time-dependent Schr\"odinger equation can be written as
\begin{equation}
	\Psi_E(x,t)=\sum_n \phi_n(x) e^{-iE_n t},
\end{equation}
which is labeled by the eigenenergy $E$ and in which $E_n=E+n\hbar\omega$.
For $x<0$ or $x>0$, the potential in Eq.~(\ref{ham}) is constant and $\phi_n(x)$ can be written as a superposition of plane waves.
For electrons originating in the left lead ($x<0$), with incident energy $E$, we can write
\begin{equation}
  \phi_n(x) = 
  \begin{cases}
      a_n \frac{e^{ik_n x}}{\sqrt{k_n}} + 
      \rho_n \frac{e^{-ik_n x}}{\sqrt{k_n}}
      & x<0    \\
       \xi_n \frac{e^{ik'_n x}}{\sqrt{k'_n}} 
       & x>0
  \end{cases}    ,
\end{equation}
where $k_n = \sqrt{2mE_n}/\hbar$ and $k'_n = \sqrt{2m(E_n- \Delta)}/\hbar$ are the wavevectors in the left and right leads, respectively, relative to the energy $E_n =  E+n\hbar\omega$.
The term with $a_n$ represents right-going incident electrons in the left lead. 
By choosing $a_n= \delta_{n,0}$ we set their energy to $E_0=E$ and we normalize the flux to 1.
The coefficient $\rho_n$
represents the left-to-left reflection amplitude, i.e. $\rho_n = {\cal S}_{LL}(E_n, E)$, for an incoming electron with energy $E$ to be scattered back at energy $E_n$. When $n\neq 0$, it corresponds to a process of inelastic reflection where the final propagating energy state $E_n>0$ has energies higher ($n>0$) or lower ($n<0$) than the initial state due to the interaction with the driven delta-like barrier. 
Analogously, the coefficient $\xi_n$ represents
the left-to-right transmission amplitude
$\xi_n={\cal S}_{RL}(E_n, E)$, non-zero only if $E_n > \Delta$.
The total transmission functions are defined according to Eq.~(\ref{eq:Tau}).
Notice that in the case of a time-independent potential, we have a finite contribution only from the term with $n=0$.

By imposing the boundary conditions for the wave-function and its derivative at $x=0$,
we obtain the following recurrence
relation for $\xi_n$
\begin{multline}\label{eq_xi}
  \Lambda_n \xi_n + 
  u_1\sqrt{\frac{k_n}{k_{n+1}}} \xi_{n+1} + 
  u_1\sqrt{\frac{k_n}{k_{n-1}}} \xi_{n-1} \\
=-ik_n \hbar^2 a_n ,
\end{multline}
where 
$\Lambda_n = u_0-i(k_n+k'_n)\hbar^2/2$
and $u_i=m W_i$, with $i=1,2$, $W_{0}$ and $W_{1}$ being introduced in Eq.~(\ref{ham}).
Equation~\eqref{eq_xi}
is characterized by two energy scales and two dimensionless parameters. The energy scales are the chemical potential $\mu$, which is related to the Fermi wavevector $k_F$ through $\mu=\hbar^2 k_F^2/(2m)$, and the step height $\Delta$.
The two dimensionless parameters are $\eta=W_0k_F/(2\mu)$, which measures the strength of the static component 
of the delta-like potential, and $v=W_1/W_0$, which is the strength of the AC component with respect to the static one.

Equation~\eqref{eq_xi} consists of an infinite set of linear equations for the coefficients $\xi_n$. 
It can be conveniently recast in a matrix form as
$\mathbf{M\cdot}\vec{\xi}=\vec{B}$,
with $\mathbf{M}$ a matrix with infinite dimensions,
and infinite dimensional vectors $\vec{\xi}=(..., \xi_1,\xi_0,\xi_{-1},...)$ and
$\vec{B}=(..., b_1,b_0,b_{-1},...)$, where $b_n=-ik_n a_n$.
For a given $\vec{B}$, which is fixed by the given incident coefficients $a_n$, such a matrix equation must be solved
for the vector $\vec{\xi}$.
Due to the infinite dimensionality of the matrix $\mathbf{M}$,
it is hard in practice to solve it by direct inversion $\vec{\xi}=\mathbf{M}^{-1}\cdot \vec{B}$. However, in the case of weak driving potentials,
it is convenient to use a perturbative approach in the small parameter
$v=W_1/W_0\ll 1$,
where one can effectively truncate the matrix $\mathbf{M}$ to a finite dimension. For example, for sufficiently small $v$, one can solve the problem considering only the energy $E$ and the first two sidebands $E_{\pm 1}=E\pm \hbar\omega$, which corresponds to the $3\times 3$ matrix 
\begin{equation}\label{matrix_M}
 \mathbf{M}=\begin{pmatrix}
 \Lambda_1 & u_1\sqrt{k_1/k_0} & 0\\
 u_1\sqrt{k_0/k_1} & \Lambda_0 & 
 u_1\sqrt{k_0/k_{-1}}\\
 0 & u_1\sqrt{k_{-1}/k_0} & \Lambda_{-1}
 \end{pmatrix}.
\end{equation}
While in this case $\mathbf{M}^{-1}$ can be computed analytically (see App.~\ref{App:3x3}), for larger dimensions it can be calculated numerically.
In particular, to reach $n$-th order accuracy, we need a $2n+1$ dimensional matrix $\mathbf{M}$.
To achieve higher accuracy, we just need to increase the dimension of $\mathbf{M}$ until convergence is attained.
In the present work, we compute our numerical results with an accuracy of at least $10^{-7}$.

\section{Results}
\label{secRes}
In this section, we present our results. We start by discussing the scenario where no step potential exists, then proceed to analyze the case where $\Delta\neq 0$. 
\subsection{Single barrier}
\label{Single}
\begin{figure}[!tbh]
\begin{center}
\includegraphics[width=0.9\columnwidth]{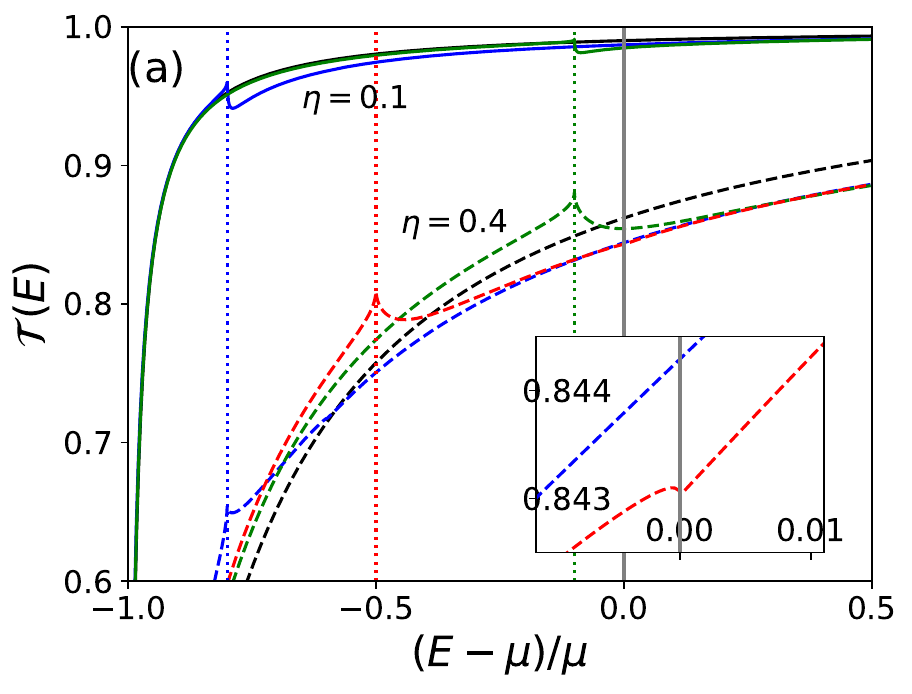} 
\includegraphics[width=0.9\columnwidth]{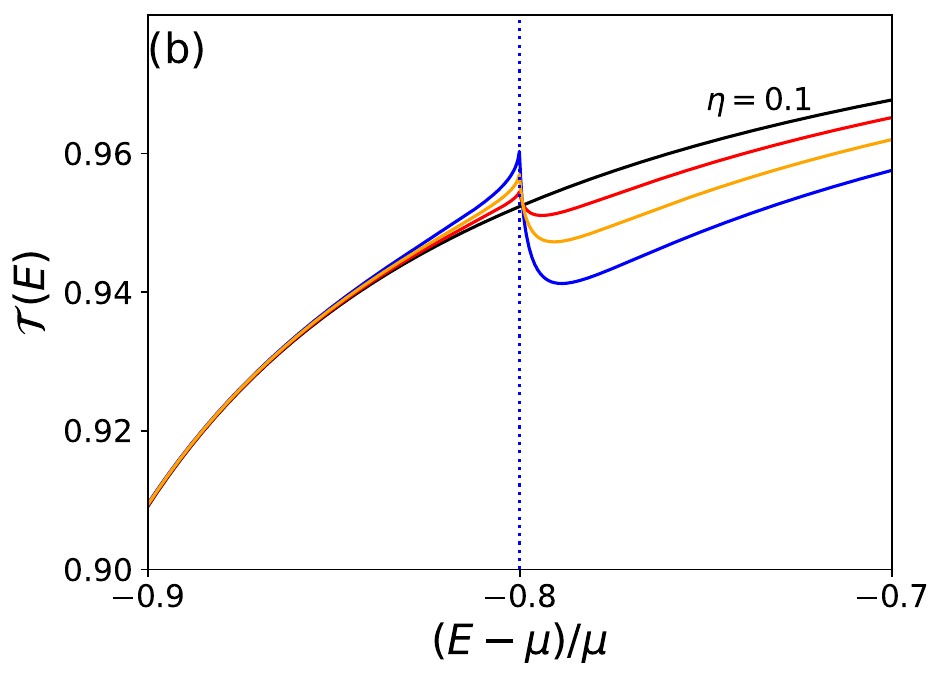}
\end{center}
\caption{\justifying Total transmission $\mathcal{T} (E)$ as the function of energy $E$.
The black curves
are for the case without AC signal, i.e. $v=0$ and serve as a reference.
(a) Comparison between the transmission $\mathcal{T} (E)$ with $\eta=0.1$ (solid) and 
$\eta=0.4$ (dashed). 
The colored curves correspond to different driving frequencies
$\hbar\omega/\mu=0.2$ (blue), $0.5$ (red) and
$0.9$ (green) and the vertical dotted lines are guides for the eyes, indicating the corresponding frequencies of the AC signal.
The thin gray vertical line indicates the chemical potential ($E=\mu$).
The inset is the zoom-in for $\eta=0.4$ around the chemical potential where $\hbar\omega/\mu=0.5$ (red) and $0.2$ (blue).
(b) Zoom-in on the kink structure around
$E-\mu=-0.8\mu$ for $\eta=0.1$.
Curves with different colors correspond to
$v =0$ (black), $0.2$ (red), $0.3$ (orange), $0.4$ (blue).}
\label{transm_without_step_art}
\end{figure}
Here, we set $\Delta=0$ in the Hamiltonian (\ref{ham}), so that the system is mirror symmetric with respect to $x=0$. This implies that the transmission amplitudes are symmetric $\mathcal{T}_{RL} (E)=\mathcal{T}_{LR}(E)\equiv\mathcal{T}(E)$.
The total transmission $\mathcal{T}(E)$, calculated using the method presented in Sec.~\ref{secMod}, is plotted in Fig.~\ref{transm_without_step_art} as a function of $E$ in units of $\mu$, for different values of
$v$, $\omega$ and $\eta$. 
The black curves are to be regarded as references corresponding to the static case, $v=0$, where the delta-like potential is time-independent. 

In Fig.~\ref{transm_without_step_art}(a) we assess the role of $\omega$ on the total transmission, for a fixed value of $v=0.4$ and two different values of $\eta=0.1$ (solid lines)$, 0.4$ (dashed lines).
The colored (solid and dashed) curves correspond to different values of the driving frequencies $\omega$ (see caption).
The plots show that the main feature produced by the AC driving is a kink structure corresponding to $E=\hbar\omega$ and its multiples, the details of which depend on the parameters.
For $\eta=0.1$, for example, the plots in panel (a) show that $\mathcal{T}(E)$ closely follows the black dashed curve up to $E=\hbar\omega$, and after the kink, it drops slightly. The effect appears to be stronger for $\hbar\omega=0.2\mu$ (blue curve). Note that in the whole range of energies considered, the value of the transmission is close to 1.
When $\eta=0.4$, for a fixed strength of the static delta-like potential, the total transmission reduces in the whole range of energies, as shown in the same panel with dashed lines. However, at the same time, the relative strength of the kinks with respect to the overall transmission profile increases.  In general, the curves corresponding to the AC driving deviate more markedly from the static case (black curve), and the kink structures become more pronounced as the driving frequency increases, as seen in the red and green curves. Notice that for energies $E\gg \hbar\omega$ all the curves with AC driving tend to the same curve, eventually converging from below to the curve for the static case.
The origin and the shape of such kink structures is discussed in detail, using an analytical perturbative approach, in App.~\ref{App:3x3}.

As already mentioned, the AC driving induces subtle higher harmonic effects that extend beyond the fundamental driving frequency. Specifically, the transmission function $\mathcal{T}(E)$ exhibits modifications not only around the primary driving energy $E=\hbar\omega$, but also at integer multiples $E=n\hbar\omega$, where $n$ represents the harmonic order. However, the effect at higher harmonics is strongly suppressed, as it is a higher order in the perturbation parameter $v$.  
In the inset of Fig.~\ref{transm_without_step_art}(a), a detailed zoom around $E=\mu$, corresponding to $E=2\hbar\omega$, reveals a particularly intriguing feature for the red dashed curve.
Namely, even a weak kink structure emerging very close to the Fermi energy could lead to measurable consequences in the Seebeck coefficient (see below).

The role of the AC driving amplitude $v$ on the total transmission is investigated in panel (b) of Fig.~\ref{transm_without_step_art}. Results are presented for $v=0$ (black curve), $0.2$ (red curve), $0.3$ (orange curve), and $0.4$ (blue curve), at a fixed frequency $\hbar\omega=0.2 \mu$ and $\eta=0.1$.
With increasing $v$, the influence of
higher harmonics of the driving frequency also emerge (not shown).

\begin{figure}[!tbh]
\centering
\includegraphics[width=0.9\columnwidth]{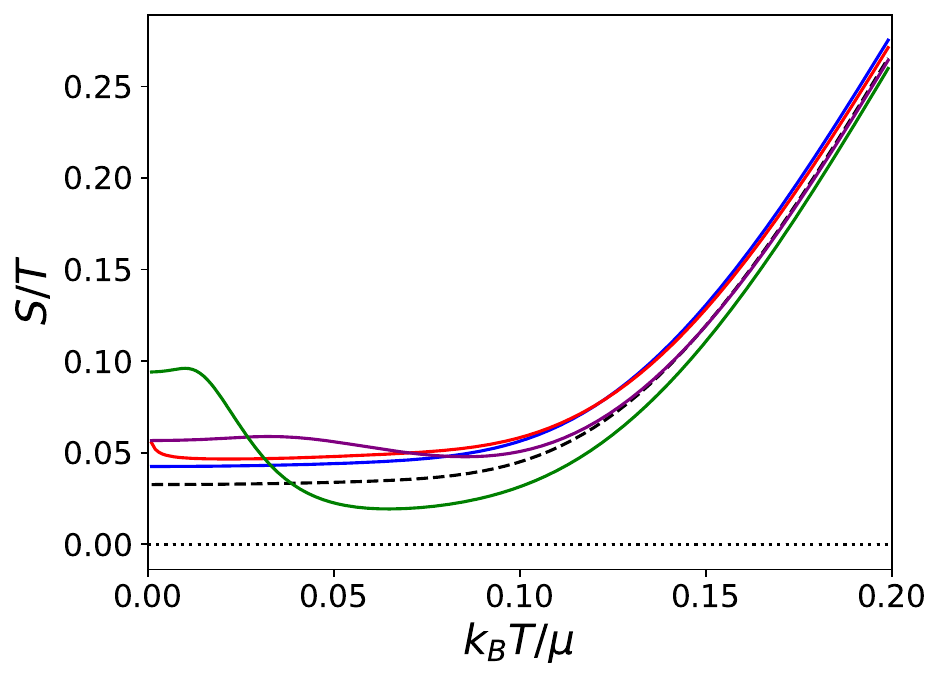}
\caption{\justifying Normalized Seebeck coefficient, $S/T\times (\mu /k_B)$, in units of $k_B/e$ as a function of the normalized temperature, $T/(\mu/k_B)$. The black dashed curve is the reference static case ($v=0$). The colored curves
correspond to $\hbar\omega/\mu= 0.2$ (blue), $0.5$ (red), $0.8$ (purple), $0.9$ (green) while $v=0.4$ and $\eta=0.1$ are fixed.}
\label{Seebeck_without_step}
\end{figure}

\subsubsection{Thermoelectric effects of AC driving}
\label{susub:thermoelectricAC}
We can now calculate how the thermoelectric properties of the system are affected by the periodic driving.
We consider the Seebeck coefficient $S$, which is derived in Sec.~\ref{FlTr} in Eq.~(\ref{eq:Seebeck}).
As discussed in Sec.~\ref{FlTr}, in the low-temperature limit ($k_BT\ll\mu$)
$S$ grows linearly with temperature $T$.
In order to emphasize the deviation from the low-temperature behavior of the Seebeck coefficient,
in Fig.~\ref{Seebeck_without_step} we plot the quantity $S/T\times (\mu/k_B)$ as a function of $k_BT/\mu$ for a fixed amplitude of oscillations $v=0.4$ and different values of frequency:
$\hbar\omega/\mu= 0.2$ (blue curve), $0.5$ (red curve), $0.8$ (purple curve), $0.9$ (green curve).
The black dashed curve represents the static situation ($v=0$), plotted for reference.

The black dashed curve
is constant for small values of $T$, as expected, but above $k_BT=0.1\mu$ starts to increases roughly linearly with temperatures (i.e.~$S\propto T^2$).
The blue curve in Fig.~\ref{Seebeck_without_step} has a similar behavior to the black dashed curve except for an enhancement at low temperatures due to the pronounced kink structure in the transmission [see the solid blue curve in Fig.~\ref{transm_without_step_art}(a)]. Specifically, the kink at $E=\hbar\omega$ leads to a corresponding suppression of $\mathcal{T} (\mu)$ of about 1\%, which appears in the denominator of Eq.~(\ref{Sapp}). Conversely, $\mathcal{T}' (\mu)$, present in the numerator of Eq.~(\ref{Sapp}), experiences a much larger increase of about 30\%. This phenomenon is illustrated in Fig.~\ref{transm_without_step_art}(a) by comparing the blue and black solid curves.
An even more pronounced effect is observed when the frequency of the AC driving increases, so that the kink structure approaches the chemical potential. 
Indeed, in Fig.~\ref{Seebeck_without_step}, we observe an increase of $S/T$ at low temperatures for increasing AC frequencies.  
Furthermore, at zero temperature, one observes a small peak in the red
curve ($\hbar\omega=0.5 \mu$). This peak is an effect of the second harmonic of the AC driving (two-photon process), which slightly changes the transmission
function around the chemical potential (i.e.~ at $E = \mu$) as discussed above. Moreover, this indicates that periodic driving may potentially enhance thermoelectricity when the driving frequency satisfies resonant conditions $\hbar\omega\simeq\mu/n$, with $n$ an integer.

We note that the green curve in Fig.~\ref{Seebeck_without_step}
shows a significantly larger value of $S/T$ at low temperatures compared to the static case (black dashed curve).
However, a strikingly different behavior emerges at higher temperatures, specifically above $k_BT\simeq 0.02\mu$, where $S/T$ decreases dramatically. This decline occurs because, within the thermal window around the chemical potential, the suppression of transmission due to the kink located at $E-\mu=-0.1\mu$ begins to take effect.
This behavior is discussed in detail in App.~\ref{AppSeebeck}.
Finally, Fig.~\ref{Seebeck_without_step} shows that in the high-temperature regime (when $\hbar\omega\lesssim  k_B T$) the AC driving only slightly corrects the static regime.

\begin{figure}[!tbh]
	\centering
\includegraphics[width=0.9\columnwidth]
{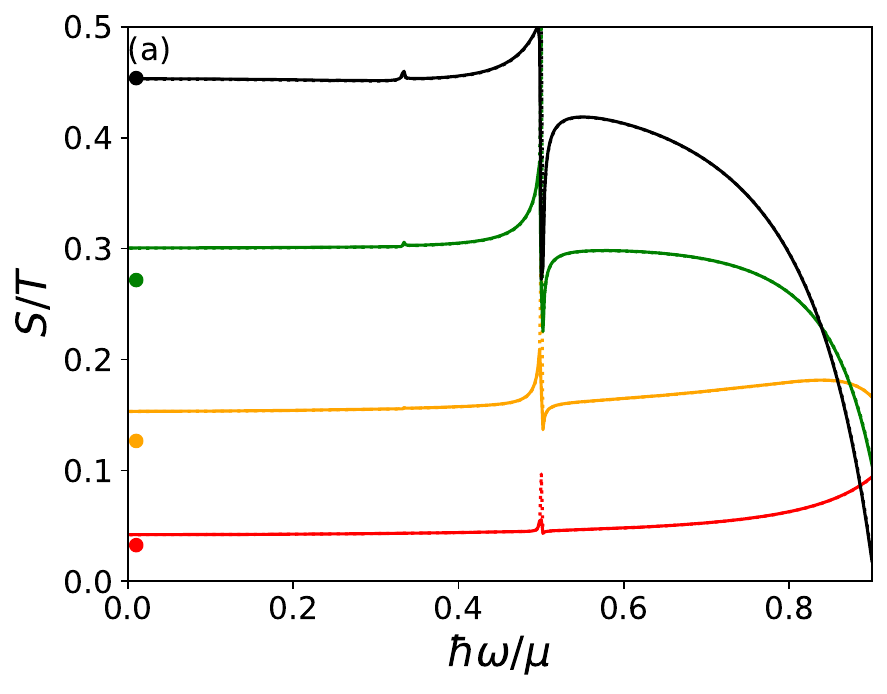}
\includegraphics[width=0.9\columnwidth]{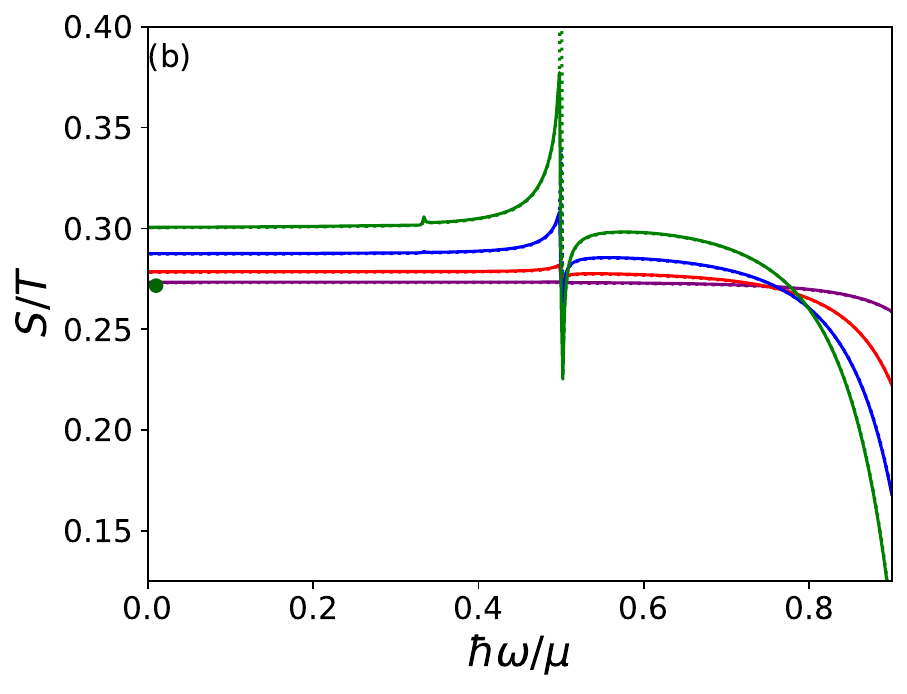}
\caption{\justifying Zero-temperature limit of the normalized Seebeck coefficient, $S/T\times (\mu /k_B) $, in units of $k_B/e$, as a function of $\hbar\omega/\mu$. (a) For different $\eta$ values:  $\eta=0.1$ (red), $0.2$ (orange), $0.3$ (green), $0.4$ (black) and a fixed strength  of oscillating delta-barrier $v=0.4$. (b) For different $v$ values: $v=0.1$ (purple), $0.2$ (red), $0.3$ (blue), $0.4$ (green) and a fixed $\eta=0.3$.
The dotted curves are $\mathcal{T}'(\mu)/\mathcal{T}(\mu)$ multiplied by $\pi^2/3$ as reference.  
The colorful full circles are the same limit for the static $v=0$ case.
}
\label{fig_curves_Seebeck_1}
\end{figure}

We can now examine how the parameters 
$v$ and $\eta$ influence the behavior of the Seebeck coefficient $S$. To this end, we present in Fig.~\ref{fig_curves_Seebeck_1} the ratio $S/T$, calculated in the zero-temperature limit, as a function of the normalized frequency $\hbar\omega/\mu$ for various values of $v$ and $\eta$. For completeness, we also report, with colored circles, the value of $S/T$  in the zero-temperature limit relative to the static case ($v=0$). It is important to note that, indeed, the small frequency limit, where $\omega\to 0$ and $v$ has a finite fixed value, does not necessarily coincide with the static case computed for $v=0$. The non-commutativity of the limits $\omega \to0$ and $v\to 0$ reflects the resonant, non-perturbative nature of the photon-assisted processes. These non-perturbative effects are the origin of the kink structures observed in the transmission functions.

In Fig.~\ref{fig_curves_Seebeck_1}(a), we fix the strength of the oscillating potential at $v=0.4$ and plot four different curves corresponding to different values of barrier strength  $\eta$ (see caption). Overall, the plot indicates that $S/T$ increases with increasing $\eta$, except for a small region of frequencies above $0.8\mu/\hbar$. For all curves, $S/T$ remains relatively constant with frequency up to approximately $\hbar\omega= \mu/2$, where a sharp feature appears. Beyond this point, the dependence on frequency transitions from an upward curvature (observed in the red and orange curves) to a downward curvature (seen in the green and black curves).
It is noteworthy that the sharp feature is quite weak for $\eta=0.1$ but becomes more pronounced as $\eta$ increases. In particular, for $\eta=0.3$ and $\eta=0.4$, this feature manifests as a peak followed by an antiresonant dip.
Notice also that an additional, though weak, feature emerges at $\hbar\omega= \mu/3$ for $\eta=0.3$ and for $\eta=0.4$.

All these features can be systematically understood by analyzing the ratio $\mathcal{T}'(\mu)/\mathcal{T}(\mu)$ which appears in the approximate expression Eq.~(\ref{Sapp}) for calculating the Seebeck coefficient $S$. Such a ratio is plotted as dotted curves in Fig.~\ref{fig_curves_Seebeck_1}(a) for the various $\eta$ values.
The solid and dotted curves overlap across the entire frequency range, except for a small region near $\hbar\omega=\mu/2$.
This shows that the pronounced sharp feature occurring around $\hbar\omega=\mu/2$
originates as a second harmonic 
effect (double-frequency effect) of the AC driving.
Indeed, as already discussed above, the latter affects the transmission function $\mathcal{T}(E)$ at energies $E=n\hbar\omega$. 
The barely visible kink occurring at $E=2\hbar\omega=\mu$ in the red curve in the inset of Fig.~\ref{transm_without_step_art}(a)
causes a large variation in $\mathcal{T}'(\mu)$, which give rise to the sharp features in $S$.
Note that the features occurring at $\hbar\omega= \mu/3$ are the effect of the third harmonic.
This discussion reveals that the kink, fundamentally linked to the significant contribution of photon-emission processes transitioning from non-propagating to propagating states at the band bottom, can influence $\mathcal{T}'(\mu)$ and consequently the thermoelectric behavior. In essence, states deep within the Fermi sea can contribute to thermoelectric performance due to photon-assisted effects.

In Fig.~\ref{fig_curves_Seebeck_1}(b), we systematically investigate the low temperature limit of the ratio $S/T$ as a function of frequency, fixing the strength of the static component of the potential to $\eta=0.3$. We plot four different curves corresponding to different values of the oscillating amplitude $v$ (see caption).
It shows an overall, though not pronounced, increase of $S/T$ with increasing $v$. More precisely, the downward bend occurring for $\hbar\omega>\mu/2$, as well as the sharp features at $\hbar\omega=\mu/2$,
become more evident as $v$ increases.
Furthermore, the effect of the third harmonic, at $\hbar\omega=\mu/3$, becomes observable only for the largest value of $v=0.4$ considered (green curve).

\begin{figure}[!tbh]
	\centering
\includegraphics[width=0.9\columnwidth]{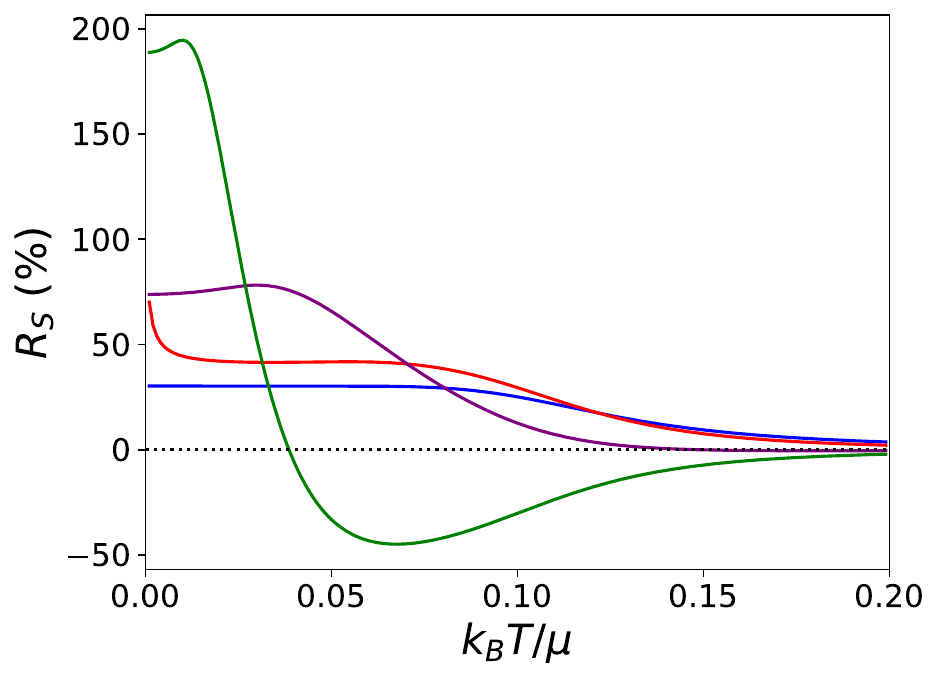}
\caption{\justifying Relative change of Seebeck coefficient $S$ (in percentage) vs $k_BT/\mu$.
The colored curves
correspond to  $\hbar\omega/\mu=0.2$ (blue), $0.5$ (red),
$0.8$ (purple), $0.9$ (green) and the other parameters are $v=0.4$ and $\eta=0.1$.}
\label{relative_without_step}
\end{figure}

To quantify the variation of $S$ in response to the AC driving, we define the relative change in the Seebeck coefficient for fixed values of $\omega$, $v$, and $\eta$ as follows
\begin{equation}
R_S=\frac{S-S_0}{S_0},
\end{equation}
where $S_0$ represents the Seebeck coefficient under static conditions (i.e., when $v=0$).
In Fig.~\ref{relative_without_step} we plot the quantity $R_S$ as a function of temperature for four values of frequency: $\hbar\omega/\mu=0.2$ (blue), $0.5$ (red),
$0.8$ (purple), $0.9$ (green) 
for the parameters which yield the largest variation, namely $v=0.4$ and $\eta=0.1$.
All curves show an enhancement of the Seebeck coefficient 
at low temperatures.
The temperature dependence of $R_{\rm S}$ is in general quite mild (see, for example, blue and purple curves in Fig.~\ref{relative_without_step}): $R_{\rm S}$ remains roughly constant in a range of temperatures and then decreases smoothly to 0.
Exceptions are the case $\hbar\omega=0.5\mu$ (red curve), where a peak appears at zero temperature, and the case $\hbar\omega=0.9\mu$ of very high frequency (green curve), where $R_{\rm S}$ decreases dramatically, reaching negative values.
Such a temperature sensitivity can be understood from the dependence on temperature of the normalized Seebeck coefficient $S$, reported in Fig.~\ref{Seebeck_without_step} and discussed above in this subsection.
In particular, the curves in Fig.~\ref{relative_without_step} exhibit the following behavior.
The blue curve ($\hbar\omega=0.2\mu$)
shows an increment of about
30\% 
in a temperature region up to 
 $T=0.1 \mu/k_B$.
As the frequency increases, the kink in the transmission approaches the chemical potential, causing the plateau in $R_S(T)$ to become higher but narrower. For the red curve ($\hbar\omega=0.5\mu$) 
the enhancement is about 40\%, in the temperature range from $0$ to $0.08  \mu/k_B$. The zero-temperature peak is due to the double-frequency effect, as mentioned before.
For the purple curve ($\hbar\omega=0.8\mu$) 
the enhancement is about 70\%, in the temperature range from $0$ to $0.05  \mu/k_B$,
while for $\hbar\omega=0.9\mu$ (green curve) the enhancement reaches 200\%, but in a smaller temperature region from $0$ to $0.02  \mu/k_B$.
Especially in the latter case, there is a trade-off between a large increase in the Seebeck coefficient and the maximum operating temperature. 

In conclusion, for systems with quite low electron density (i.e.,~small Fermi energy) the AC driving could indeed increase the Seebeck coefficient in the quantum regime  $k_BT\ll\hbar \omega\lesssim \mu$.
We expect the effect
to appear in semiconducting nanostructures, such as nanowires, with
a gate-tunable barrier~\cite{Kammhuber2017}, where thermoelectric measurements can also be performed at cryogenic temperatures~\cite{Kopp1975,Dutta2019,Svilans2018,Chen2018}.
In this case, the typical values for the Seebeck coefficient are within
reach of experimental parameters.
For example, reading from Fig.~\ref{Seebeck_without_step}, for $v = 0.4$, $\eta = 0.1$ and $\hbar\omega/\mu = 0.5$ (red curve), one obtains $S/T=0.07k_B^2/(e\mu)$ for $k_B T /\mu=0.1$. By choosing $\mu=1$ meV, one finds $S=0.6$ $\mu V/$K at frequencies of hundreds of GHz.

\subsection{Single barrier with a step}
In this section, we consider the scenario with a step function potential and set $\Delta > 0$ in the Hamiltonian given by Eq.~(\ref{ham}).
Since $\Delta$ breaks spatial inversion symmetry, the transmission functions from left to right $\mathcal{T}_{RL}$ and from right to left
$\mathcal{T}_{LR}$ are not necessarily equal when an AC driving is applied.

\begin{figure}
\centering
\includegraphics[width=0.9\columnwidth]{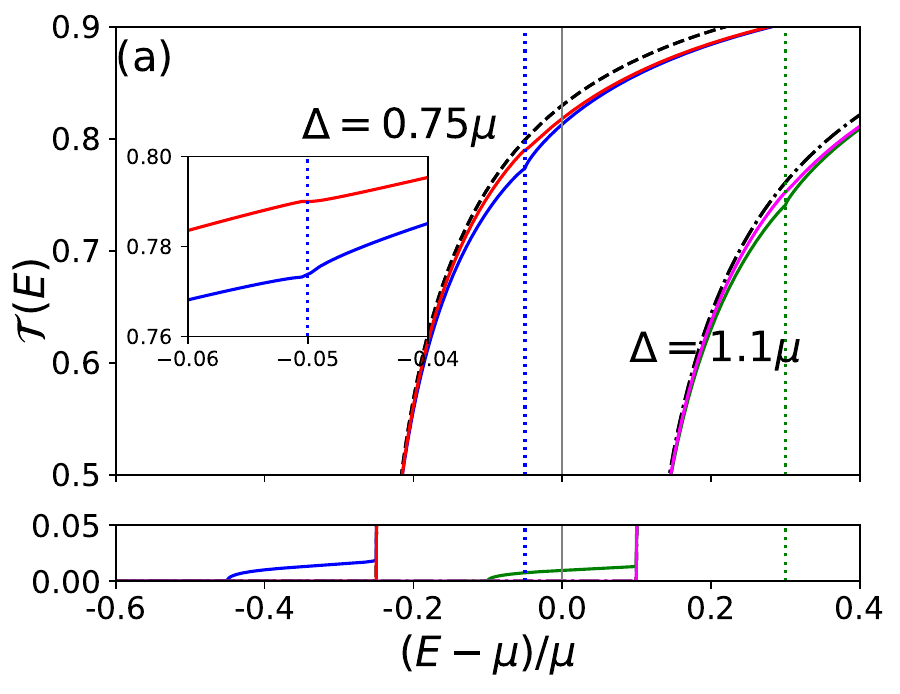}
\includegraphics[width=0.9\columnwidth]{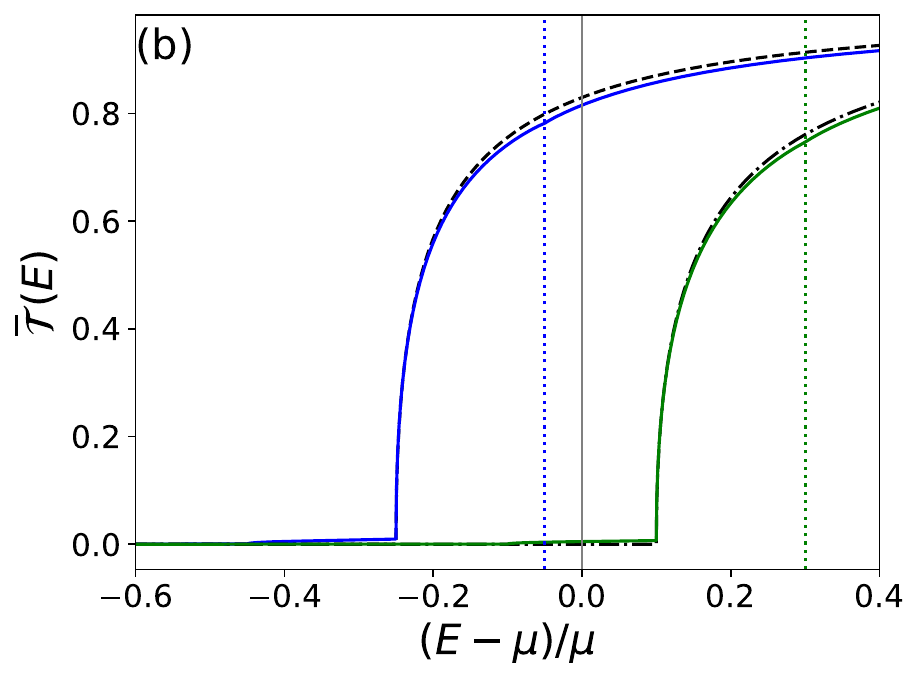}
\caption{
\justifying 
Single barrier with a step potential: energy dependence of transmission functions. 
(a) Blue and green curves are for the left-to-right transmission $\mathcal{T}_{RL}$, while red and magenta curves are for right-to-left transmission $\mathcal{T}_{LR}$.
One set of curves (red and blue) refers to the step height  $\Delta=0.75\mu$, and the other (green and magenta) refers to $\Delta=1.1\mu$. The inset is the zoom-in of the kink structures around $E=0.95\mu$.
(b) Blue (green) curve is the average transmission $\bar{\mathcal{T}}= (\mathcal{T}_{RL} + \mathcal{T}_{LR})/2$ for $\Delta=0.75\mu$ ($\Delta=1.1\mu$).
For both panels, black dashed curves refer to the static case
with $\Delta =0.75\mu$, while black dash-dotted curves correspond to the static case with $\Delta =1.1\mu$. 
Vertical dotted lines are guides
for the eyes, indicating the corresponding frequencies of
the AC signal.
Vertical thin gray lines indicate the chemical potential, i.e.
$E=\mu$.
Other parameters are: $\eta=0.2$, $v=0.4$ and $\hbar\omega=0.2\mu$.
}
\label{fig_step_trms}
\end{figure}

In Fig.~\ref{fig_step_trms}(a), we show these transmissions as functions of energy $E$ for a fixed frequency $\hbar\omega=0.2\mu$. The plot is divided into two transmission ranges: 0.5 to 0.9 (top panel) and 0 to 0.05 (bottom panel), allowing for a clearer visualization of features present at low transmission values.
Two sets of curves illustrate two distinct scenarios with different step heights: for $\Delta = 0.75\mu$ (blue and red curves), the chemical potential is above the step height, while for $\Delta = 1.1\mu$ (green and magenta curves), the chemical potential is below the step height.
The blue and green curves represent the transmissions $\mathcal{T}_{\rm RL}$, while the red and magenta curves correspond to $\mathcal{T}_{\rm LR}$. Dashed black curves depict the static case, included for reference, for which the equality $\mathcal{T}_{RL} = \mathcal{T}_{LR}$ 
holds.
In the static case, as illustrated in Fig.~\ref{fig_step_trms}(a), transmission is zero for energies below the step ($E < \Delta$), since the electronic states on the right lead are evanescent (do not propagate).
In contrast, AC driving leads to a non-zero transmission $\mathcal{T}_{RL}$ even for energies below $\Delta$, as evident from the green and blue curves in the figure.
This is due to the fact that
an incident electron from the left lead, possessing energy $E < \Delta$, can overcome the potential step and transfer to the right lead into a higher energy state ($E+n\hbar\omega > \Delta$) by absorbing 
$n$ energy quanta through AC driving.
On the other hand, for $E < \Delta$, the red and magenta curves (right-to-left transmission, $\mathcal{T}_{LR}$) exhibit behavior analogous to the static case: they are zero because incident states from the right lead cannot propagate below the step.
Figure~\ref{fig_step_trms}(a) also shows that for energies above the step, all AC-driven curves lie slightly below their static counterparts (dashed lines). 

Similar to the findings in Section~\ref{Single}, albeit less pronounced, additional kink structures appear at $E=\Delta+\hbar\omega$ (marked by the vertical dotted lines). To enhance visibility of these features, in Fig.~\ref{fig_step_trms}(a), we include an inset showing a zoom-in of the red and blue curves ($\Delta=0.75\mu$).
Note that the kinks for the two transmissions in the blue and red curves go in opposite directions (one goes down and the other goes up, respectively), so that when summed together, they may even compensate.
This is shown in Figure~\ref{fig_step_trms}(b) where we plot the averaged transmissions 
$\bar{\mathcal{T}}= (\mathcal{T}_{RL} + \mathcal{T}_{LR})/2$.
Note that here the range of values of transmissions is different from the one used in panel (a) and ranges between 0 and 0.9.
Finally, both Figs.~\ref{fig_step_trms}(a) and (b) 
show that there is no qualitative difference between the group of curves for $\Delta=1.1\mu$ and $\Delta=0.75\mu$, apart from the obvious shift in energy.

Before proceeding, we note that because of the difference between $\mathcal{T}_{RL}(\epsilon )$ and $\mathcal{T}_{LR}(\epsilon )$, a non-zero pumping current can flow through the system even in the absence of voltage or thermal biases (i.e.~$\Delta V=0$ and $\Delta T=0$).
Indeed, according to Eqs.~(\ref{chcu}) and (\ref{je}), in this case, the charge current is given by
\begin{equation}\label{pump_charge_cur}
J_e=J_e^P=\frac{e}{h}\int_{-\infty}^{\infty} dE
\left[  \mathcal{T}_{LR}(E ) 
    -
\mathcal{T}_{RL}(E ) \right]f(E) .
\end{equation}
More details are reported in Appendix~\ref{Appump}.
We can now study the thermoelectric properties of the system and calculate the coefficient $L_{\rm eh}$ defined in Eq.~(\ref{Leh}).
Here we do not consider the Seebeck coefficient $S$ because the charge current, and hence the coefficient $L_{\rm ee}$, is strongly suppressed when the chemical potential $\mu$ is lower than the height of the step $\Delta$. Indeed, in this situation, the behavior of $S$ is totally dominated by $L_{\rm ee}$, see Eq.~(\ref{eq:Seebeck}).

\begin{figure}[!tbh]
\centering
\begin{subfigure}{0.4\textwidth}
\includegraphics[width=\columnwidth]{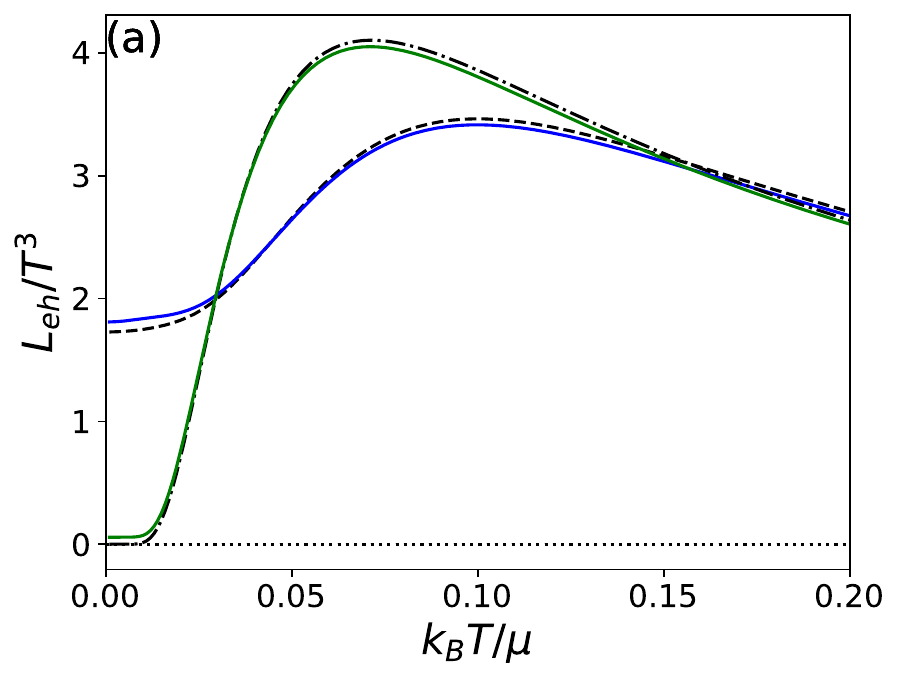}
\end{subfigure}
\hfill
\begin{subfigure}{0.4\textwidth}
\includegraphics[width=\columnwidth]{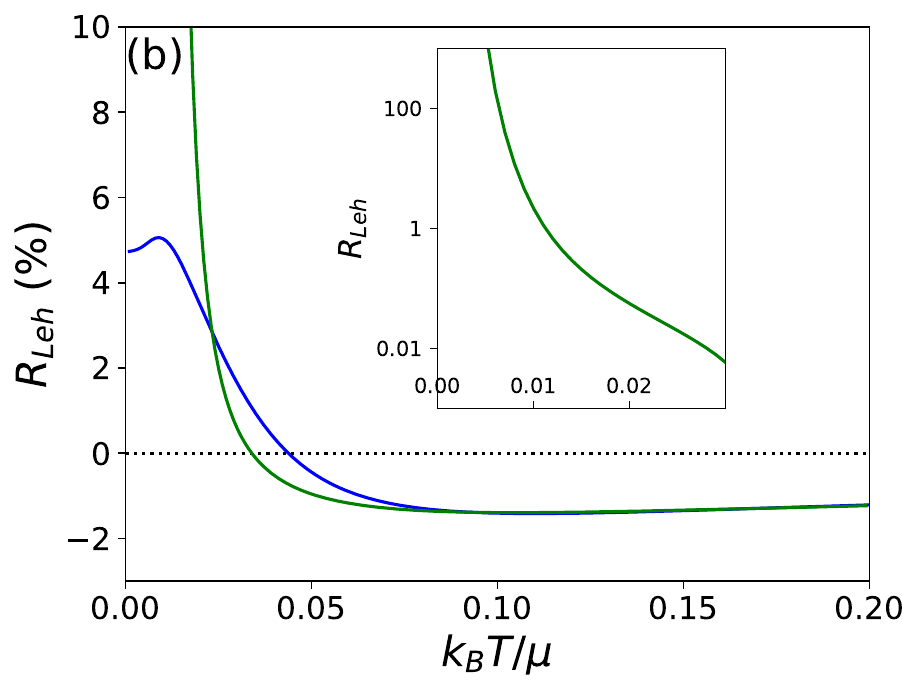}
\end{subfigure}
\caption{\justifying 
Single barrier with a step potential: (a)~normalized thermoelectric coefficient $L_{\rm eh}/T^3$, in units of $\mu/(ek_B^2)$,
and (b)~its relative change $R_{L_{\rm eh}}$, as functions of the normalized temperature $k_BT/\mu$.
Blue curves are relative to $\Delta=0.75\mu$, green curves are relative to $\Delta=1.1\mu$, and dashed black curves refer to the static cases.
Other parameters are: $\eta=0.2$, $v=0.4$ and $\hbar\omega=0.2\mu$. 
In the main plot of panel (b) $R_{L_{\rm eh}}$ is expressed in percentage, while the inset shows the low-temperature $R_{L_{\rm eh}}$ (actual ratio) in the logarithm scale for $\Delta=1.1\mu$ only.
}
\label{fig_Leh_step}
\end{figure}

Since in the static situation $L_{\rm eh}$ scales as $T^3$ in the low-temperature regime, see Sec.~\ref{FlTr}, in Fig.~\ref{fig_Leh_step} we plot the ratio $L_{\rm eh}/T^3$ as a function of the rescaled temperature $k_BT/\mu$ for the two values of potential step height: blue curve for $\Delta= 0.75\mu$ and green curve for $\Delta= 1.1\mu$. The dashed black curves refer for the static case.
The main difference between the two is that, at $T=0$, one takes a finite value ($\Delta= 0.75\mu$), while the other becomes zero ($\Delta= 1.1\mu$).
Both curves exhibit a broad peak at intermediate values of temperatures (between 0.05 and 0.1 $k_BT/\mu$).
Figure~\ref{fig_Leh_step}(a) shows that the AC driving, in both cases, causes $L_{\rm eh}$ to slightly increase at small $T$ and slightly decrease at higher $T$.
In both cases, however, the two curves do not deviate significantly from the static case.
This behavior is discussed in detail in App.~\ref{AppSeebeck}.

To quantify the variation of $L_{\rm eh}$ due to the AC driving, with respect to the static situation, we proceed similarly to what we did before, introducing the relative change
\begin{equation}\label{R_Leh}
R_{L_{\rm eh}}=\frac{L_{\rm eh}-L_{\rm eh}^{(0)}}{L_{\rm eh}^{(0)}},
\end{equation}
where $L_{\rm eh}^{(0)}$ denotes the static thermoelectric coefficient, i.e., the value of $L_{\rm eh}$ when $v = 0$.
The quantity $R_{L_{\rm eh}}$ is plotted in Fig.~\ref{fig_Leh_step}(b) as a function of the normalized temperature for the two different values of $\Delta$, for fixed values of $\omega$, $v$, and $\eta$.
The blue and green curves represent the cases with the step value $\Delta=0.75\mu$
and $\Delta=1.1\mu$, respectively.
Both curves show an enhancement of $L_{\rm eh}$ in the low-temperature regime. 
The blue curve has a maximum of about 5\% increase, while the green curve
shows a much larger increase, since it diverges for $T$ going to zero.
Such a divergence is entirely due to the vanishing of the static coefficient $L_{\rm eh}^{(0)}$.
Indeed, the photon-assisted effect of the driving enhances the thermoelectric coefficient $L_{\rm eh}$ in a range of temperatures where transport is otherwise suppressed by the presence of the potential step.
The inset shows a logarithmic scale zoom-in in a range of small temperatures.

We stress that, from an experimental point of view, the changes due to the driving in $L_{\rm eh}$ are observable.
Indeed, in a realistic nanowire setup, see for example Ref.~\cite{Kammhuber2017}, a chemical potential $\mu=1-10$ meV and a gap $\Delta=1.1\mu$ are achievable with gate-tunable devices. In this case, a change of $L_{\rm eh}$ on the order of $0.1 (e/h) k_B^2/\mu$ [read out from the green curve in Fig.~\ref{fig_Leh_step}(a) at a small temperature] corresponds to detecting a current, see Eq.~(\ref{je}), $J_{\rm e}~\sim~2-20$ pA at $T=1$ K and $\Delta T= 0.1$ K, which is within experimental reach.

\section{Conclusions}
\label{secConclusion}
In the present work, we investigated the effects of an AC driving on the thermoelectric properties in the framework of Floquet theory.
Two simple models are studied in one-dimensional space, both including a delta-function like potential whose strength oscillates in time (the AC driving).
In the first model, the system exhibits left-to-right symmetry, which can be viewed as a homogeneous, highly doped 1D semiconductor. In contrast, the second model breaks this symmetry by introducing a step function potential, which may describe an inhomogeneous semiconducting interface.
We assume that the electron density in the system is very low, such that the AC driving frequency can be as large as fractions of the Fermi energy and/or comparable with the semiconducting gap.

Our results for the first model showed that driving could strongly enhance the Seebeck coefficient, especially at low temperatures, by increasing the strength of the AC component of the delta-like potential with respect to the static one.
In particular, we found that the relative change of the Seebeck coefficient with respect to the static condition could reach 200\% at large frequencies, where the corresponding energy significantly exceeds the thermal energy, i.e.~in the quantum regime.
In the second model, we focused on the thermoelectric Onsager coefficient, finding that its relative change with respect to the static case could be significantly enhanced at low temperatures, especially when the chemical potential is within the semiconductor's gap, where the current is negligible in static conditions.

In the second model, where left-to-right symmetry is broken, we note that additional current contributions arise from a non-adiabatic pumping effect. This effect has the potential to partially obscure the enhancement of thermoelectricity observed in the symmetric case. Despite this, by isolating the terms linear in the temperature difference, it is still possible to single out the impact of photon-assisted processes on thermoelectricity. 
The second model not only describes inhomogeneous semiconductors but also offers a potentially useful analogy for phenomenology at a normal metal/superconductor interface, where the superconducting gap arises from the pairing of electrons. Such an analogy appears relevant for the field of superconducting quantum technologies, where superconductors are operated at cryogenic temperatures in the high-frequency (quantum) domain.

To conclude, our results demonstrate that external driving (i.e.~coherent photon-assisted processes) can be used to tune and enhance the thermoelectric capabilities of low-electron-density nanodevices.
For the future, it would be very interesting to investigate the effect of high-frequency driving on a resonant system, where thermoelectric properties are already excellent in the static situation.

\section{Acknowledgements}
We acknowledge the Royal Society through the International Exchanges between the UK and Italy (Grants No. IEC/R2/212041). F.T. and A.B. acknowledge the 
MUR-PRIN 2022—Grant No. 2022B9P8LN-(PE3)-Project NEThEQS “Non-equilibrium coherent thermal effects in quantum systems” in PNRR Mission 4-Component 2-Investment 1.1 “Fondo per il Programma Nazionale di Ricerca e Progetti di Rilevante Interesse Nazionale (PRIN)” funded by the European Union-Next Generation EU and CNR project QTHERMONANO. A.B. acknowledges discussions with F. Giazotto and A. Jordan.

\begin{appendix}
\section{Methods to calculate the transmission coefficients within the Floquet theory}
\label{App:3x3}
As mentioned in Sec.~\ref{secMod} of the main text, 
to obtain the vector of transmission coefficients $\vxi$ in the Floquet theory, one has to treat the infinite linear equations $\mathbf{M}\cdot\vec{\xi}=\vec{B}$,
representing the set of Eqs.~(\ref{eq_xi}),
with $\mathbf{M}$ an infinite-sized known matrix and $\vec{B}$ a known vector related to the amplitude of the incident wave.
The matrix $\mathbf{M}$ can be decomposed into $\mathbf{M}=\mathbf{\Lambda}+u_1 \mathbf{M}_1$, in which the matrix $\mathbf{\Lambda}$ is a diagonal 
matrix [with elements $(...,\Lambda_{1},\Lambda_{0},\Lambda_{-1},...)$],
and the matrix $\mathbf{M}_1$ is zero on the diagonal.
We can use two approximate methods to treat the problem.
The first one is perturbative: $\vec{\xi}$ can be calculated
order by order by assuming that
$u_1$ is a small parameter.
The second method is
to truncate the matrix $\mathbf{M}$ into a  finite dimension
and then solve the truncated matrix equation (finite-dimensional).
In practice, the latter approach must be implemented numerically. 
Both methods are accurate to a certain order
(power) in $v=u_1/u_0$, and their numerical results clearly converge when the order increases.

Let us first consider the perturbation method.
By expressing the matrix $\mathbf{M}$ as
$\MM=\mathbf{\Lambda}+ u_1 \mathbf{M}_1$, 
and considering $u_1 \MM_1$ as a perturbation,
one can solve $\MM\cdot\vec{\xi}=\vec{B}$ perturbatively to find
\begin{equation}
\vxi=\vxi^{(0)}+ u_1\vxi^{(1)} + u_1^2\vxi^{(2)} + \mathcal{O} (u_1^3) ,
\end{equation}
in which $\vxi^{(0)}=\MLambda^{-1}\cdot \vec{B}$ and
$\vxi^{(l+1)}=-(\MLambda^{-1}\cdot \MM_1)\cdot\vxi^{(l)}$ for $l\ge 1$.

If the incident component $a_n$  is given by
$a_n= \delta_{n,0}$,
the leading orders $\vxi^{(0)}$ and
$\vxi^{(1)}$ are 
\begin{equation}
\vxi^{(0)}=  \xi^{(0)}_0=-ik_0/\Lambda_0
 \end{equation}
and
\begin{equation}
\vxi^{(1)}=
\begin{bmatrix}
     \xi^{(1)}_1 \\           
    \xi^{(1)}_0 \\
    \xi^{(1)}_{-1}
 \end{bmatrix} =
ik_0/\Lambda_0
 \begin{bmatrix}
     \sqrt{k_1/k_0}/\Lambda_1 \\
     0 \\
     \sqrt{k_{-1}/k_0}/\Lambda_{-1}
 \end{bmatrix} .
 \label{eq:A2}
\end{equation}
For both infinite vectors $\vxi^{(0)}$ and  $\vxi^{(1)}$, all the other components are zero
(therefore omitted).
We also calculate $\vxi^{(2)}$, obtaining
\begin{equation}\label{xi2}
\vxi^{(2)}= -ik_0/\Lambda_0
 \begin{bmatrix}
     \sqrt{k_2/k_0}/(\Lambda_2\Lambda_1)\\
     0 \\
      (\Lambda_{1}^{-1}+\Lambda_{-1}^{-1})/\Lambda_0\\
     0 \\
     \sqrt{k_{-2}/k_0}/(\Lambda_{-2}\Lambda_{-1})
 \end{bmatrix} .
\end{equation}

Recalling that the transmission probability is defined as
$\mathcal{T}(E)=\sum'_n |\xi_n|^2$,
in order to compute the $\mathcal{T}(E)$ up to second order in $u_1$ we need to compute only the terms derived above, i.e., $\vxi^{(0)}$, $\vxi^{(1)}$ and $\vxi^{(2)}$.

\begin{figure}[!tbh]
	\centering
\includegraphics[width=0.9\columnwidth]{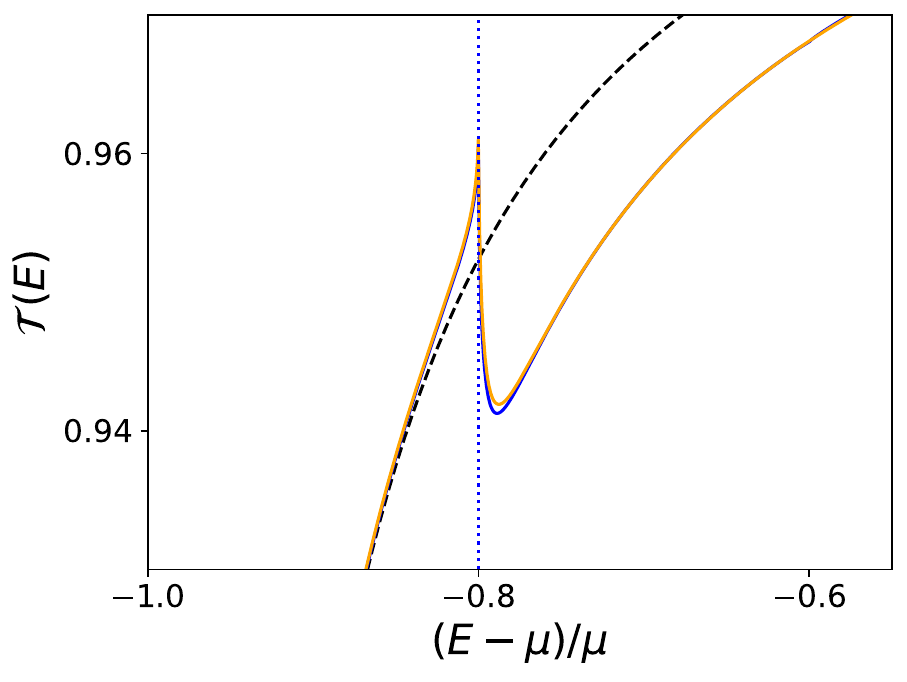}
\caption{\justifying Transmission function $\mathcal{T}(E)$: comparison between the perturbation result (orange curve) and the truncation result (blue curve),
with the parameters $\eta=0.1$ and $v=0.4$.
As for the truncation,
we keep 5 positive modes and 5 negative modes, i.e.
$n=-5,-4,...,5$, with a total of 11 components.
The black dashed curve is the static case.
The vertical dotted line indicates the energy of the AC driving with $\hbar\omega=0.2\mu$.}
\label{fig_append_pert}
\end{figure}

We now turn our attention to a comparison of the perturbation and truncation methods. The energy dependence of the transmission function $\mathcal{T}(E)$ is depicted in Fig.~\ref{fig_append_pert}, where the dashed black curve is a reference corresponding to the static case, the orange curve is calculated using the perturbative approach, and the blue curve is obtained from the truncation method accounting for 11 modes. The plot shows a very good agreement between the two methods.

Interestingly, the occurrence of the kink structure in the transmission function around $E=\hbar\omega$ (especially the sharp drop on the right-hand side of the kink,
i.e. for $E>\hbar\omega$) can be explained using the perturbative method as follows. 
If the amplitude of the AC signal is small, i.e. $v=u_1/u_0 \ll 1$,
we can express the transmission function as 
$\mathcal{T}(E)= |\xi^{(0)}_0 + u_1^2 \xi^{(2)}_0|^2 + u_1^2 |\xi^{(1)}_1|^2
+ u_1^2 |\xi^{(1)}_{-1}|^2$, in which the terms with higher powers than $u_1^2$ are
neglected.
Inserting the results obtained above, we have
$\mathcal{T}(E)=  (k_0^2 /|\Lambda_0|^2)\mathcal{W}(E) $
with $\mathcal{W}=\mathcal{W}_0 +\mathcal{W}_{1} +\mathcal{W}_{-1}$
in which 
\begin{align}\label{W_modulo_square}
 \mathcal{W}_0 = \left| 1-i\phi_0 + \frac{v^2}{1-i\phi_1} + \frac{v^2}{1-i\phi_{-1}}\right|^2 / |1-i\phi_0 |^2,
\end{align}
$\mathcal{W}_{1} =v^2 (\phi_1/\phi_0)
/|1-i\phi_1 |^2$ and
$\mathcal{W}_{-1} =v^2 (\phi_{-1}/\phi_0)
/|1-i\phi_{-1} |^2$,
and $\phi_i = k_i/u_0$ ($i=-1,0,1$).

Considering an energy $E$ close to $\hbar\omega$ in such a way that 
$E-\hbar\omega=\epsilon$ 
(with $\epsilon$ small and positive),
we can expand $\mathcal{W}$ in 
powers of $\delta = \sqrt{2m\epsilon}/u_0$
finding
\begin{align}\label{W_modulo_square_2}
\mathcal{W}(E) = \mathcal{W}(\hbar\omega)
+ v^2\mathcal{N}\delta + \mathcal{O}(\delta^2) ,
\end{align}
with $\mathcal{N}=(1-\varphi_0^2)/(\varphi_0(1+\varphi_0^2))$.
In Eq.~(\ref{W_modulo_square_2}), terms with higher power than 
$v^2$ are omitted, and
$\varphi_0$ is
the limit value of $ \phi_0$ when $\epsilon\to 0$.
By comparing with the notations introduced in Sec.~\ref{secMod},
one finds that $\varphi_0^2 = (\hbar\omega/E_F)/\eta^2$.
For the case with $\eta=0.1$ and $\hbar\omega=0.2 \mu$,
we have $\mathcal{N}<0$, and therefore $\mathcal{T}(E)$
decreases with $E$, when $E$ is a little bigger than $\hbar\omega$.
Furthermore, $\delta$ has a strong dependence on $E$, when $E$ is near $\hbar\omega$ from the right side, which explains the sharp peak structure
at $E=\hbar\omega$.

\section{Pumped current}
\label{Appump}
In the presence of the step function potential
(say $\Delta > 0$), the AC driving makes
the left-to-right transmission
function different from the right-to-left one, as shown in Fig. \ref{fig_step_trms}.
According to Eq.~(\ref{chcu}), the charge  current is given by
\begin{equation}
J_e=\frac{e}{h}\int_{-\infty}^{\infty} dE 
\left[  \mathcal{T}_{RL}(E ) 
f_L(E )\right.     
- \left.\mathcal{T}_{LR}(E ) f_R(E)\right] ,
\end{equation}
which reduces to Eq.~(\ref{pump_charge_cur}), the pumped current $J_e^P$,
in the absence of voltage and thermal biases.

\begin{figure}[!tbh]
\centering
\includegraphics[width=0.9\columnwidth]{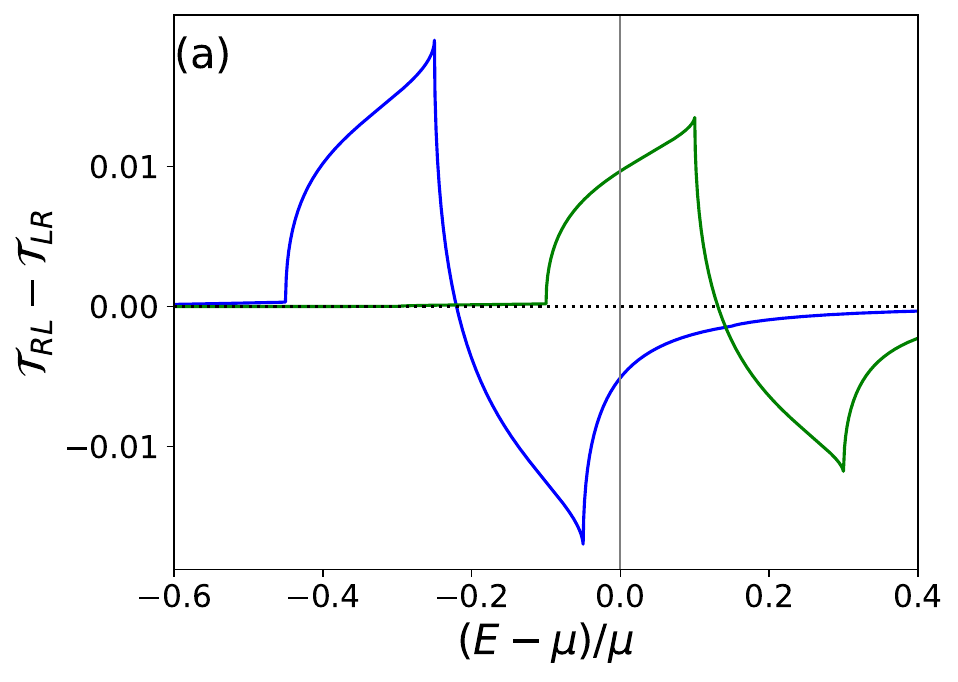}
\includegraphics[width=\columnwidth]{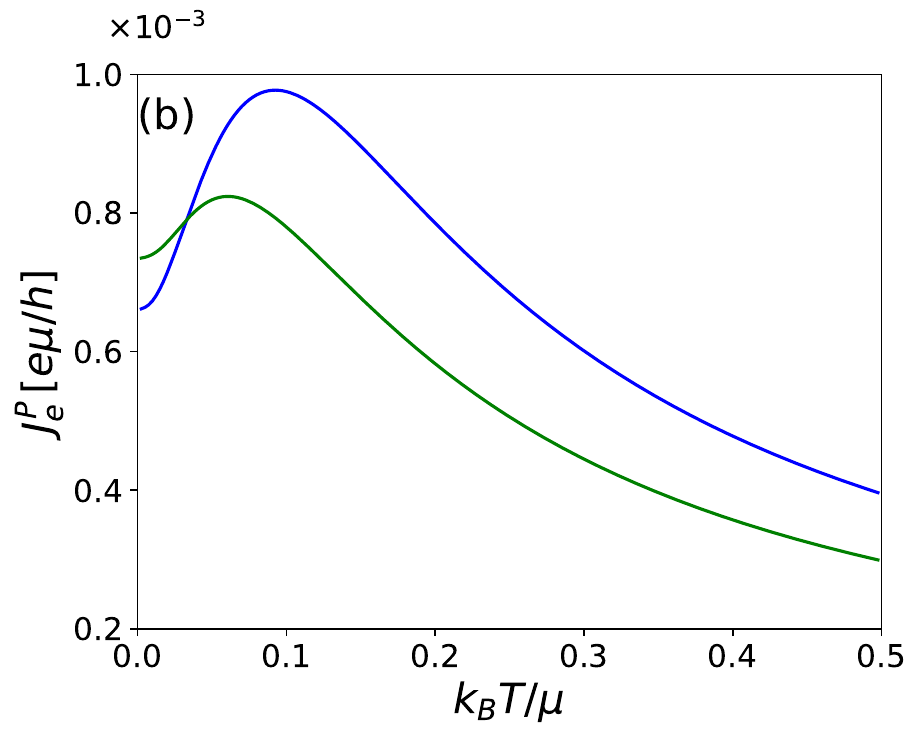}
\caption{\justifying (a) Difference between $\mathcal{T}_{RL}$ and $\mathcal{T}_{LR}$ 
as a function of energy. The vertical dotted line refers to $E=\mu$.
(b) Pumped current in units of $e\mu/h$ as a function of temperature $T$.
The blue and green curves correspond to $\Delta=0.75\mu$ and 
$\Delta=1.1\mu$, respectively.}
\label{pump}
\end{figure}

Fig.~\ref{pump}(a) shows the difference between 
$\mathcal{T}_{RL}$ and $\mathcal{T}_{LR}$ for the two values of $\Delta$.
One finds that for $E<\Delta$,
$\mathcal{T}_{RL} > \mathcal{T}_{LR}$, while for $E>\Delta$,
$\mathcal{T}_{RL} < \mathcal{T}_{LR}$.
Notably, sharp kinks emerge at $E=\Delta$ and at $E=\Delta+\hbar\omega$.
For energies far from $\Delta$
the difference  between $\mathcal{T}_{RL}$ and $\mathcal{T}_{LR}$ goes to zero.
The pumped current as a function of temperature
is shown in Fig.~\ref{pump}(b).

\section{Detailed behavior of the Seebeck coefficient}
\label{AppSeebeck}
\subsection{Single barrier}
\begin{figure}[!tbh]
\centering
\includegraphics[width=0.9\columnwidth]{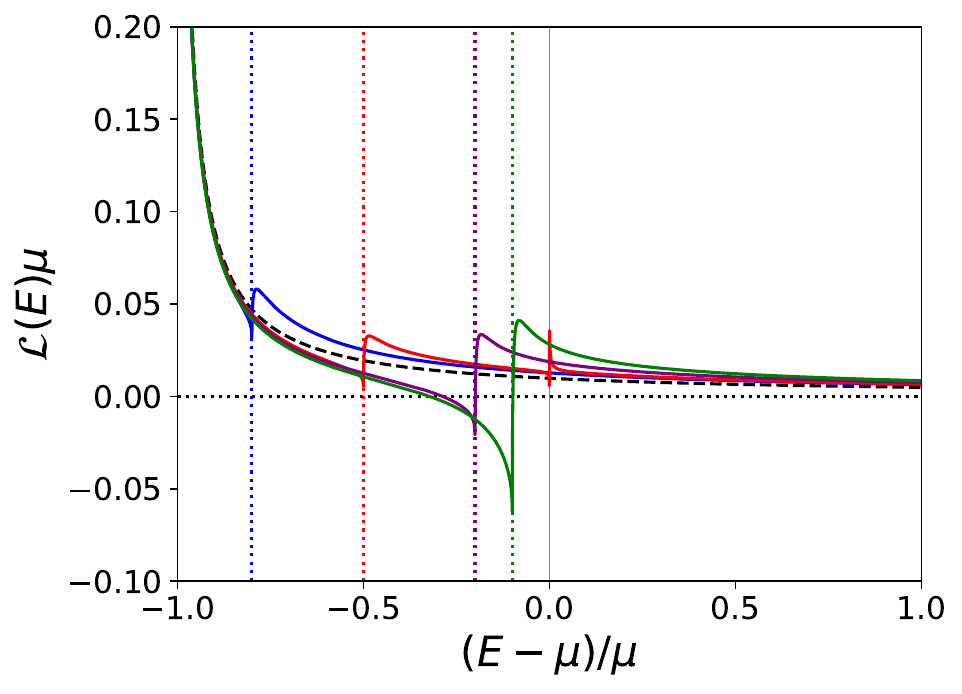}
\caption{\justifying Auxiliary function $\mathcal{L} (E)\mu$ defined in the text.
The colored curves
correspond to $\hbar\omega/\mu=0.2$ (blue), $0.5$ (red), 0.8 (purple) and  $0.9$ (green), while $v=0.4$ and $\eta=0.1$ are fixed. The black dashed curve represents the static case.}
\label{fig_Lagrange}
\end{figure}

In order to understand the behavior of the Seebeck coefficient in Fig.~\ref{Seebeck_without_step},
we need to focus on the numerator of Eq.~(\ref{eq:Seebeck}), since the denominator, which represents the conductance, remains virtually temperature-independent in the range of temperatures considered (varying by only about 1\%).
The numerator of the Seebeck coefficient, proportional to $L_{\rm eh}$, can be expressed as
$\int dE (E-\mu)^2 \mathcal{L} (E) f'(E)$,
where we introduce the auxiliary function
\begin{equation}
\label{ellfunction}
    \mathcal{L}(E)=\frac{\mathcal{T}(E)-\mathcal{T}(\mu)}{E-\mu} ,
\end{equation}
plotted in Fig.~\ref{fig_Lagrange}.
It is important to note that $\mathcal{L}(E)$ exhibits a sharp feature at $E=\hbar\omega$, where it can take negative values, as illustrated by the green curve. The integrand is influenced by temperature through the derivative $f'(E)$, which is non-zero in a range of energies on the order of $k_BT$ around $\mu$
\footnote{$f'(E)$ is a delta function at zero temperature, 
and broadens when the temperature increases.}.

Consequently, when the temperature is sufficiently high for $f'(E)$ to encompass this sharp feature, the integrand reduces in value. 
This leads to a decrease in
$S/T$ starting from $k_BT\simeq 0.02\mu$ for the green curve.
In the case $\hbar\omega=0.8\mu$ (purple curve), the effect of the kink structure on $S$ is much weaker, but still causes a decrease of $S/T$ starting from $k_BT\simeq 0.04\mu$.
In the other cases involving the red ($\hbar\omega=0.5\mu$) and blue ($\hbar\omega=0.2\mu$) curves, the kink structure at $E=\hbar\omega$ gradually moves away from the chemical potential.
This leads to a behavior of $\mathcal{L}(E)$ that closely resembles the static case (see Fig.~\ref{fig_Lagrange}). As a result, the Seebeck coefficient behaves similarly to the static scenario,  following the black dashed curve
and increasing at higher temperatures.

\subsection{Single barrier with a step}
The overall behavior of all the curves in Fig.~\ref{fig_Leh_step}(a) can be understood from the static situation as follows.
At zero temperature, the quantity  
$L_{\rm eh}/T^3$ takes the limiting value
 $(\pi^2/3)(e/h)k_B^2\bar{\mathcal{T}}' (\mu)$~\cite{Benenti_2017}.
 \begin{figure}[!tbh]
\centering
\includegraphics[width=0.9\columnwidth]{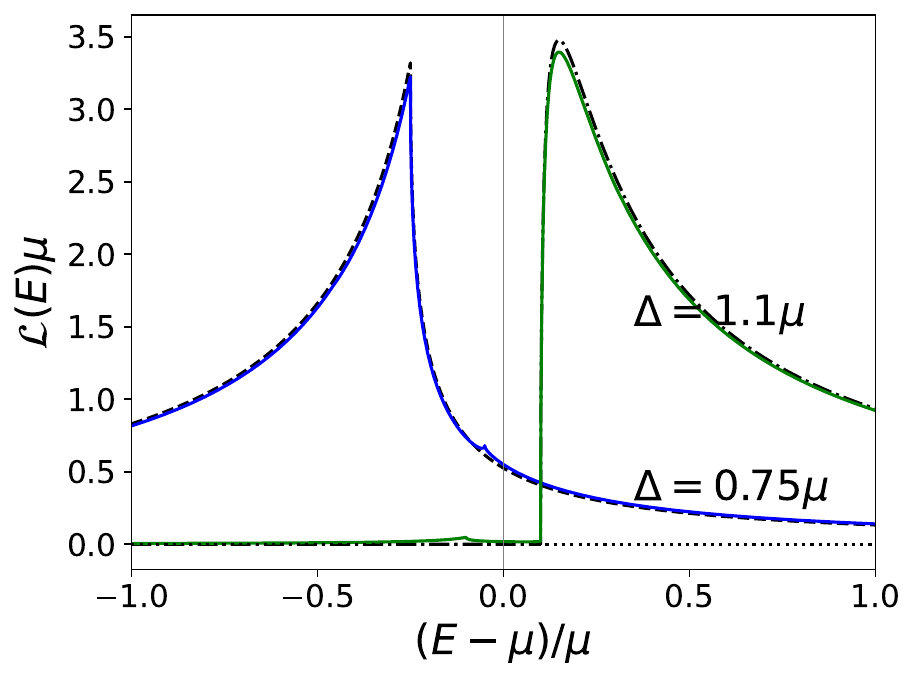}
\caption{\justifying 
Auxiliary function $\mathcal{L} (E)$ defined in the text.
The blue and green curves
correspond to $\Delta=0.75\mu$ and $\Delta=1.1\mu$, respectively, while other parameters are: $\omega=0.2\mu$, $\eta=0.1$ and $v=0.4$. Black dashed curves represent the static case. The value of the parameters $\Delta$ used is indicated in the figure.}
\label{fig_Lagrange_step}
\end{figure}

In order to understand the temperature dependence of $L_{\rm eh}/T^3$,
we follow the approach of the previous subsection and introduce the auxiliary 
function $\mathcal{L}(E)$, defined by Eq.~(\ref{ellfunction}), 
in which the transmission function $\mathcal{T} (E)$ is replaced by the 
averaged transmission $\bar{\mathcal{T}} (E)$.
Fig.~\ref{fig_Lagrange_step} shows the plot of the function $\mathcal{L}(E)$ with a non-zero potential step. The value of $\mathcal{L}(\mu)$ is very close to the value of the derivative of $\bar{\mathcal{T}} (E)$ at the chemical potential, i.e. $\bar{\mathcal{T}}' (\mu)$.
A key feature of the curves of $\mathcal{L}(E)$ in Fig.~\ref{fig_Lagrange_step} is the peaks at $E=\Delta$. 
As mentioned in the previous subsection, at $T=0$, the function
$f'(E)$ is a delta function centered at $E=\mu$. When temperature increases, the function $f'(E)$ broadens
and the integration in Eq.~(\ref{Leh}) averages the function $(E-\mu)^2\mathcal{L}(E)$ over an interval around $E=\mu$. 
If the broadening includes the peak
structure of $\mathcal{L}(E)$, the value of 
 $L_{\rm eh}/T^3$ increases.
However, at higher temperatures, when the averaging interval extends beyond the peak, the contribution from the peak diminishes, leading to a decrease in $L_{\rm eh}/T^3$.

We now consider the effect of the AC driving.
Figure~\ref{fig_Lagrange_step} shows that the blue and green curves are slightly higher than their dashed counterparts in the vicinity of $E=\mu$.
This reflects in a larger values of $L_{\rm eh}$ for small temperatures, as shown in Fig.~\ref{fig_Leh_step}(a).
However, beyond the peak, i.e.~in the energy range $E<\Delta$ for the blue curve and  $E>\Delta$ for the green curve, the AC driving reduces the value of $\mathcal{L}(E)$.
This explains why in Fig.~\ref{fig_Leh_step}(a) the AC driving reduces the value of $L_{\rm eh}$ in the high-temperature regime.

\end{appendix}
\bibliography{revised-bibliography}
\end{document}